\documentclass[showkeys,showpacs,superscriptaddress,nofootinbib]{revtex4-2}
\usepackage{amsmath}
\usepackage{amssymb}
\usepackage{graphicx}
\usepackage{bm}
\usepackage{float}
\usepackage{yfonts}
\usepackage{ulem,xcolor}
\usepackage{color,soul}

\begin{document}
	
\title{
	Charge gap in SU(3) Yang-Mills-plus-nonlinear-spinor-field theory
}
	
\author{
	Vladimir Dzhunushaliev
}
\email{v.dzhunushaliev@gmail.com}
	
\affiliation{
	Department of Theoretical and Nuclear Physics,  Al-Farabi Kazakh National University, Almaty 050040, Kazakhstan
}
\affiliation{
	Institute of Experimental and Theoretical Physics,  Al-Farabi Kazakh National University, Almaty 050040, Kazakhstan
}

\affiliation{
	Academician J.~Jeenbaev Institute of Physics of the NAS of the Kyrgyz Republic, 265 a, Chui Street, Bishkek 720071, Kyrgyzstan
}

\affiliation{
	Laboratory for Theoretical Cosmology, International Centre of Gravity and Cosmos,
	Tomsk State University of Control Systems and Radioelectronics (TUSUR),
	Tomsk 634050, Russia
}	
\author{Vladimir Folomeev}
\email{vfolomeev@mail.ru}

\affiliation{
	Institute of Experimental and Theoretical Physics,  Al-Farabi Kazakh National University, Almaty 050040, Kazakhstan
}	

\affiliation{
	Academician J.~Jeenbaev Institute of Physics of the NAS of the Kyrgyz Republic, 265 a, Chui Street, Bishkek 720071, Kyrgyzstan
}

\affiliation{
Laboratory for Theoretical Cosmology, International Centre of Gravity and Cosmos,
Tomsk State University of Control Systems and Radioelectronics (TUSUR),
Tomsk 634050, Russia
}
		
\author{Daulet Berkimbayev}
	
\affiliation{
	Department of Theoretical and Nuclear Physics,  Al-Farabi Kazakh National University, Almaty 050040, Kazakhstan
}
\date{\today}
	
\begin{abstract}
Particlelike solutions in SU(3) gauge Yang-Mills theory with color magnetic and electric fields sourced by a nonlinear spinor field are obtained. The asymptotic behavior of these fields is studied. It is shown that the electric field exhibits the Coulomb asymptotic behavior, and one of the color components of the magnetic field  behaves asymptotically as the field of a magnetic dipole. This allows one to determine the corresponding charge and magnetic moment. It is shown that profiles of the color charge and magnetic moment have global minima, which may be called charge and magnetic moment gaps. The relationship between the total energy of the system and the color charge is obtained. The physical reason for the appearance of the mass, charge, and magnetic moment gaps in the theory under consideration is discussed.
\end{abstract}
	
	
	
\date{\today}
	
\maketitle
	
\section{Introduction}

In quantum chromodynamics, there is a well-known problem which states that in a non-Abelian quantum Yang-Mills theory there should exist a minimum value of the mass, called a mass gap.
In this connection, one might wonder whether minimum values of other physical quantities in quantum field theories do exist. For instance, one may consider the question of the existence of a charge gap.
It would be logical to assume that the presence of a gap for any physical quantity should be related to the
presence of a mass gap. If such a relationship does exist, one might expect that if in some theory there is
a global minimum in the energy spectrum of particlelike solutions sourced by a charged field (a spinor field or
a complex scalar field), for such a spectrum, there would exist a global minimum for a charge of such particlelike solutions.

In Ref.~\cite{Ayyar:2017qii}, it is shown how weak coupling Feynman diagrams can contain the information of a nonperturbative mass gap in an asymptotically free theory. In Ref~\cite{Burikham:2017bkn}, it is demonstrated that if one takes a QCD glueball as a massive spin-2 state to be mixed with a massless graviton, then strong gravity will predict a mass gap of any object composed of particles that couple to the glueball. In Ref.~\cite{Faedo:2017fbv}, it is shown that holographic duals of gauge theories with a mass gap also exhibit confinement. In Ref.~\cite{Kojo:2012js}, the behavior of a mass gap in the presence of a magnetic field is studied, and it is shown that the quark mass stays around $\Lambda_{\text{QCD}}$ or less, provided that it is mainly generated by the nonperturbative part of the gluon exchange. However, that still leaves open the question of whether there exist minimum values of other physical quantities.

Consistent with this, we will demonstrate here the existence of a global minimum for a charge, i.e., the presence of a charge gap. A crucial role in our investigations will be played by nonlinear spinor fields.  In the 1950's, the nonlinear Dirac equation was employed by  W.~Heisenberg in constructing a model of an electron~\cite{heis}. In turn, various modifications of this equation were used within the Nambu-Jona-Lasinio model (for a review, see Refs.~\cite{Volkov:2005kw,Buballa:2003qv}).

In the present study we generalize the equations obtained in Refs.~\cite{Dzhunushaliev:2020qwf,Dzhunushaliev:2021apa}
[where a non-Abelian magnetic field belongs to the group  SU(2), which can be regarded as a subgroup  $\text{SU(2)} \subset \text{SU(3)}$]
by introducing a $t$-component of a non-Abelian gauge potential belonging to the subgroup
$\text{U(1)} \subset \text{SU(3)}$.
The presence of such a component results in the appearance of a color electric field sourced by a nonlinear spinor field $\psi$. As was demonstrated in Refs.~\cite{Dzhunushaliev:2020qwf,Dzhunushaliev:2021apa}, in this theory, the energy spectrum of particlelike solutions has a global minimum~-- a mass gap. It was shown in Ref.~\cite{Dzhunushaliev:2021apa} that the dominant contribution to the energy density is given by the term $\psi^\dagger \psi$ that is obviously related to the density of color charge as well. This fact enables us to assume that for a family of regular solutions belonging to the subgroup
$\text{SU(2)} \times \text{U(1)} \subset \text{SU(3)}$  (with a color electric field included) obtained in the present paper, there should exist a global minimum of a color charge~-- a charge gap.

\section{General equations and  field \textit{Ans\"{a}tze}}
\label{YM_Dirac_scalar}

In this section we closely follow Ref.~\cite{Dzhunushaliev:2020qwf}. The Lagrangian describing a system consisting of a non-Abelian SU(3) field $A^a_\mu$ interacting with nonlinear spinor field $\psi$ can be taken in the form
\begin{equation}
\begin{split}
	\mathcal L = & - \frac{1}{4} F^a_{\mu \nu} F^{a \mu \nu}
	+ i \hbar c \bar \psi \gamma^\mu D_\mu \psi  -
	m_f c^2 \bar \psi \psi+
	\frac{l_0^2}{2} \hbar c \left( \bar \psi \psi \right)^2.
\label{1_10}
\end{split}
\end{equation}
Here $m_f$ is the mass of the spinor field;
$
D_\mu = \partial_\mu - i \frac{g}{2} \lambda^a
A^a_\mu
$ is the gauge-covariant derivative, where $g$ is the coupling constant and $\lambda^a$ are the SU(3) generators (the Gell-Mann matrices);
$
	F^a_{\mu \nu} = \partial_\mu A^a_\nu - \partial_\nu A^a_\mu +
	g f_{a b c} A^b_\mu A^c_\nu
$ is the field strength tensor for the SU(3) field, where $f_{a b c}$ are the SU(3) structure constants; $l_0$ is a constant; $\gamma^\mu$ are the Dirac matrices in the standard representation; $a,b,c=1, \cdots, 8$ are color indices and $\mu, \nu = 0, 1, 2, 3$ are spacetime indices.

Using the Lagrangian~\eqref{1_10}, one can find the corresponding field equations,
\begin{eqnarray}
	D_\nu F^{a \mu \nu} &=&
	\frac{g \hbar c}{2}
	\bar \psi \gamma^\mu \lambda^a \psi =  j^{a \mu} ,
\label{1_20}\\
	i \hbar \gamma^\mu D_\mu \psi  - m_f c \psi + l_0^2 \hbar \psi
	\left(
		\bar \psi \psi
	\right)&=& 0 .
\label{1_25}
\end{eqnarray}
Here $j^{a \mu}$ is a color current created by the spinor field $\psi$. Particlelike solutions are sought in the form of
\begin{align}
 A^a_i = & \frac{1}{g} \left[ 1 - f(r) \right]
	\begin{pmatrix}
		0 & \phantom{-}\sin \varphi &  \sin \theta \cos \theta \cos \varphi \\
		0 & -\cos \varphi &   \sin \theta \cos \theta \sin \varphi \\
		0 & 0 & - \sin^2 \theta
	\end{pmatrix} , \quad
		i = r, \theta, \varphi  \text{ (in spherical coordinates)},
		\quad a = 1,2,3 ,
\label{1_30}\\
	A^8_t = & \frac{\chi(r)}{g r},
\label{1_40}
\end{align}
where \eqref{1_30} is  the standard \textit{Ansatz} used in describing the 't~Hooft-Polyakov monopole and 
written in spherical coordinates (in cartesian coordinates, one can find the corresponding expression, for example, in the textbook~\cite{Cheng}); here, the index $i$ labels the columns and the index  $a$ labels the rows. In turn, the \textit{Ansatz} for the spinor field is taken in the form 
\begin{equation}
	\psi_i = \frac{e^{-i \frac{E t}{\hbar}}}{g r \sqrt{2}}
	\begin{pmatrix}
		0 & u &	i v \sin \theta e^{- i \varphi}	&	- i v \cos \theta \\
		-u & 0 &	- i v \cos \theta										&	- i v \sin \theta e^{i \varphi} \\
		0 & 0 &	0																							&	0
	\end{pmatrix}, 
\label{1_50}
\end{equation}
where the index $i$ labels the isospinors. This is the SU(3) generalization of the SU(2) \textit{Ansatz} given in Refs.~\cite{Li:1982gf,Li:1985gf}. Here $E/\hbar$ is the spinor frequency and the functions $u$ and $v$ depend on the radial coordinate $r$ only.

For the \textit{Ans\"{a}tze} \eqref{1_30} and \eqref{1_40},
we have the following nonvanishing components of electric and magnetic fields:
\begin{align}
	E^8_r = & \frac{\chi - r \chi'}{g r^2} , \quad E^8_\theta = E^8_\varphi = 0, 
\label{1_60}\\
	H^1_i = & \frac{1}{g}
	\left\lbrace
		\frac{\sin \theta \cos \varphi}{r^2} \left(1 - f^2\right),
		-\cos \theta \cos \varphi \, f',
		\sin \theta \sin \varphi \, f'
	\right\rbrace ,
\label{1_70}\\
	H^2_i = & \frac{1}{g}
	\left\lbrace
		\frac{\sin \theta \sin \varphi}{r^2} \left(1 - f^2\right),
		-\cos \theta \sin \varphi \, f',
		- \sin \theta \cos \varphi \, f'
	\right\rbrace ,
\label{1_80}\\	
	H^3_i = & \frac{1}{g}
	\left\lbrace
		\frac{\cos \theta}{r^2} \left(1 - f^2\right),
		\sin \theta \, f',
		0
	\right\rbrace ,
\label{1_90}
\end{align}
where the spatial index $i = r, \theta, \varphi$.

Substituting the expressions~\eqref{1_30}-\eqref{1_50} in the field equations~\eqref{1_20} and \eqref{1_25}, one can obtain
equations for the unknown functions $f, u$, $v$, and $\chi$:
\begin{align}
	- f^{\prime \prime} + \frac{f \left( f^2 - 1 \right) }{x^2} =&
	- \tilde g^2 \frac{\tilde u \tilde v}{x}  ,
\label{1_100}\\
	- \frac{\chi^{\prime \prime}}{x} = & - \tilde g^2
	\frac{\tilde u^2 + \tilde v^2}{2 \sqrt{3} \, x^2} ,
\label{1_110}\\
	\tilde v' + \frac{f \tilde v}{x} =& \tilde u \left(
	- 1 + \tilde E + \frac{\chi}{2 \sqrt{3} \, x}
	+ \frac{\tilde u^2 - \tilde v^2}{x^2}
	\right) ,
\label{1_120}\\
	\tilde u' - \frac{f \tilde u}{x} =& \tilde v \left(
	- 1 - \tilde E - \frac{\chi}{2 \sqrt{3} \, x}
	+ \frac{\tilde u^2 - \tilde v^2}{x^2}
	\right),
\label{1_130}
\end{align}
written in terms of the following dimensionless variables:
$x = r/\lambda_c$,
$
	\tilde u=u\sqrt{l_0^2/\lambda_c g^2},
	\tilde v = v\sqrt{l_0^2/\lambda_c g^2},
	\tilde E = E/(m_f c^2),
	\tilde g^2 = \left( \bar g \lambda_c/ l_0 \right)^2
$,
where $\lambda_c= \hbar / (m_f c)$ is the Compton wavelength and ${\bar g}^2 = g^2 \hbar c$ is a dimensionless coupling constant. The prime denotes differentiation with respect to $x$. Notice that Eq.~\eqref{1_110} is in fact the Maxwell equation for the color electric field $\vec E^8$,
$
	\mathbf{\nabla} \cdot \vec E^8 = \rho_c ,
$
with $\vec E^8$ taken from~ Eq.~\eqref{1_60}, and  the density of the color electric charge 
is defined according to the expression for the current density  $j^{a \mu}$ from the Yang-Mills equation~\eqref{1_20}:
\begin{equation*}
		\rho_c \equiv j^{8 t} = - \frac{c \hbar}{2 \sqrt{3}} \frac{u^2 + v^2}{g r^2}
	= - \frac{1}{g \lambda_c^3}
	\tilde g^2 \frac{\tilde u^2 + \tilde v^2}{2 \sqrt{3} \, x^2}. 
\end{equation*}
The total energy density of the system is
\begin{equation}
	\tilde \epsilon =
	\frac{1}{\tilde g^2}
	\left[
		\frac{{f'}^2}{ x^2} +
		\frac{\left( f^2 - 1 \right)^2}{2 x^4}
		+ \frac{
			\left( \chi - x \chi'\right)^2
		}{2 x^4}
	\right] +
	\left[
		\tilde E \frac{\tilde u^2 + \tilde v^2}{x^2} +
		\frac{\left(\tilde u^2 - \tilde v^2 \right)^2}{2 x^4}
	\right] =
	 \tilde \epsilon_{\text{YM}} + \tilde \epsilon_s,
\label{1_140}
\end{equation}
where the expressions in the first square brackets correspond to the energy density 
of color magnetic and electric fields $\tilde \epsilon_{\text{YM}}$  
and the expressions in the second square brackets describe the energy density 
of the nonlinear spinor field $\tilde \epsilon_s$. Then the total energy of the system under consideration is calculated as
\begin{equation}
	\tilde W_t  \equiv \frac{\lambda_c g^2}{\tilde g^2} W_t= \tilde W_{\text{YM}}
	+ \tilde W_{s} =
	4 \pi
	\int\limits_0^\infty x^2 \tilde \epsilon d x .
\label{1_150}
\end{equation}

\section{Numerical solutions, electric charge and magnetic moment
}
\begin{figure}[h]
\includegraphics[width=1\linewidth]{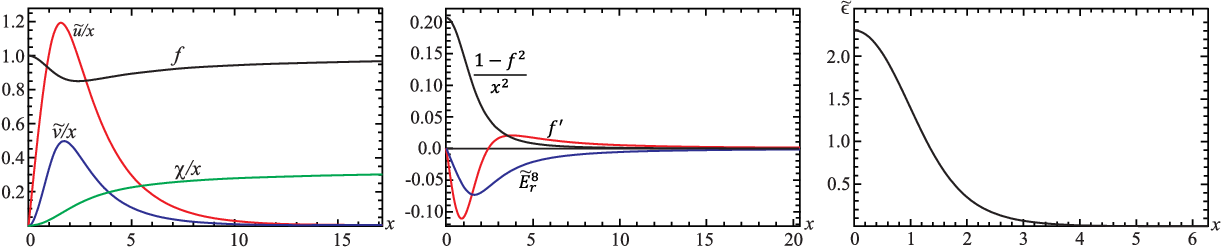}
\vspace{-0.5cm}
\caption{
Left panel: the gauge potentials $f, \chi$ and the spinor functions $u,v$.
Middle panel: the radial component of the color electric field  $\tilde E^8_r$, the radial part $(1 - f^2)/x^2$
of the radial components of the color magnetic field  $H^{1,2,3}_r$, and the radial part $f^\prime$
of the tangential components of the color magnetic field $H^{1,2,3}_{\theta, \varphi}$.
Right panel: the profile of the energy density \eqref{1_140}. For all panels,
$\tilde g = 0.7$, $\tilde E = 0.8$, $f_2 = -0.206395$, $\tilde u_1 = 1.1971179$.
}
\label{fig_fields}
\end{figure}

A solution of Eqs.~\eqref{1_100}-\eqref{1_130} in the vicinity of  the center $x = 0$ can be obtained
from the expansion of the functions in a Taylor series:
\begin{equation}
\begin{split}
	f = 1 + \frac{f_2}{2} x^2 + \ldots ,\quad
	\tilde u = \tilde u_1 x + \frac{\tilde u_3}{3!} x^3 + \ldots ,\quad
	\tilde v = \tilde v_0 + \frac{\tilde v_2}{2} x^2 + \frac{\tilde v_4}{4!} x^4 	+ \ldots,
	\quad
	\chi = \chi_1 x + \frac{\chi_3}{3!} x^3
	+ \ldots
\end{split}
\nonumber
\end{equation}
Substitution of these expansions in the field equations yields the following values of the expansion coefficients:
\begin{equation}
	\tilde v_0 = 0, \quad
    \tilde v_2 = \frac{2}{3} \tilde u_1 \left(
	\tilde E - 1 + \tilde u_1^2
	\right) , \quad 
   \chi_1 = 0, \quad
    \chi_3 = \frac{\tilde g^2}{2 \sqrt{3}} \tilde u_1^2 ,
\nonumber
\end{equation}
whereas the parameters $\tilde u_1$ and $f_2$ are free. The functions $f$ and $v$ are even, while $v$ and $\chi$ are odd with respect to the origin of coordinates; this is seen from the corresponding symmetries of Eqs.~\eqref{1_100}-\eqref{1_130} under the substitution $x \rightarrow -x$. The coefficients $\tilde u_3$ and $\tilde v_4$ can also be obtained upon substitution of the corresponding Taylor series in  Eqs.~\eqref{1_100}-\eqref{1_130}, but for assigning the boundary conditions it is sufficient to know only the coefficients $ f_2, \tilde u_1, \tilde v_2$, and $\chi_3$; for this reason, we did not calculate the coefficients $\tilde u_3$ and $\tilde v_4$. Solutions possessing a finite total energy do  exist  not for all values of 
$\tilde u_1$ and $f_2$, which are eigenvalues of the problem. To find them, we solved Eqs.~\eqref{1_100}-\eqref{1_130} numerically using the shooting method to obtain regular solutions with finite energy. The typical solutions are depicted in  Fig.~\ref{fig_fields}.

\begin{figure}[h]
\begin{minipage}[t]{.48\linewidth}
	\begin{center}
		\includegraphics[width=1\linewidth]{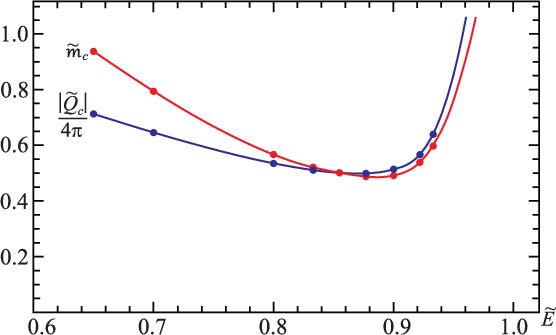}
	\end{center}
\vspace{-0.5cm}
\caption{The dependence of the modulus of the color charge $\left|\tilde Q_c \right|$
and of the color magnetic dipole moment $\tilde{\mathfrak{m}}_c$ on the parameter $\tilde E$.
}
\label{charge_mgn_moment}
\end{minipage} \hfill
\begin{minipage}[t]{.48\linewidth}
	\begin{center}
		\includegraphics[width=1\linewidth]{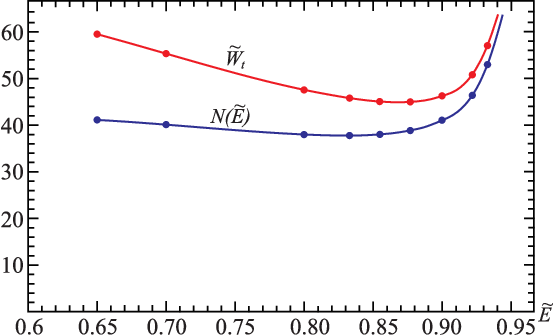}
	\end{center}
\vspace{-0.5cm}
\caption{The energy spectrum of the solutions $\tilde W_t$
and the normalization integral
$
	N\left(\tilde E \right)  = 4 \pi \tilde E \int \limits_0^\infty
	\left( \tilde u^2 + \tilde v^2\right) dx
$ which is proportional to the absolute value of the color charge from Eq.~\eqref{3_47}.
}
\label{energy_spectrum}
\end{minipage} \hfill
\end{figure}

Using the results of numerical calculations,
one can show that asymptotically (as $x\to \infty$) the functions $f(x)$ and $\chi(x)$ behave as
\begin{align}
	f \approx & 1 - \frac{\tilde{\mathfrak{m}}_c}{x} ,
\label{3_30}\\
	\tilde A^8_t\equiv g\lambda_c A^8_t = \frac{\chi}{x} \approx & \chi_\infty + \frac{\tilde Q_c}{x},
\label{3_40}
\end{align}
where $\tilde{\mathfrak{m}}_c$ and $\tilde Q_c$ are constants whose physical meaning is defined below.

As it follows from Eq.~\eqref{3_40}, the color electric field has a Coulomb-law fall off at large distances,
as  expected due to the fact that Eq.~\eqref{1_110} describes in practice an Abelian electric field.
Then the electric field strength has the following asymptotic behavior:
\begin{equation}
\tilde	E^8_r \approx \frac{\tilde Q_c}{x^2}.
\label{3_45}
\end{equation}
Thus the parameter $\tilde  Q_c$ appearing in the asymptotic expressions  \eqref{3_40} and \eqref{3_45}
is a dimensionless color electric charge for the U(1) field. To find its numerical values, one can construct profiles of the distribution of $\tilde E^8_r x^2$ as a function of the radial coordinate $x$, and then determine the asymptotic value of this distribution. Alternatively, one can calculate the integral of the density of color charge
 \begin{equation}
	\tilde Q_c \equiv g Q_c	= - \frac{2 \pi \tilde g^2 }{\sqrt{3}}
	\int \limits_0^\infty
		\left( \tilde u^2 + \tilde v^2\right) dx .
\label{3_47}
\end{equation}
It is seen that the color charge is negative; in order to obtain a positive value of the charge, it is necessary to take an opposite sign of the  \textit{Ansatz} \eqref{1_40} for the potential $A^8_t$.

We have checked that the values of the charge  $\tilde Q_c$ determined according to Eqs.~\eqref{3_45} and \eqref{3_47} are the same for $\tilde E = 0.7$. Next, we have studied the dependence of the modulus of the color charge $\left|\tilde Q_c \right|$ on the parameter $\tilde E$ (shown in Fig.~\ref{charge_mgn_moment}). It is seen from this figure that the color charge $\left|\tilde  Q_c \right|$ has a global minimum, which may be called a charge gap. This gap is present in SU(3) Yang-Mills theory under consideration with a source of a color electric field taken in the form of a nonlinear spinor field.

Fig.~\ref{energy_spectrum} shows the distributions of the energy [given by Eq.~\eqref{1_150}] and of the normalization integral
as functions of the spinor frequency $\tilde E$. In turn,
using the analytic expression for the charge \eqref{3_47}, one can obtain the following relationship
between the total energy \eqref{1_150} and the color charge~\eqref{3_47}:
\begin{equation}
	\tilde W_t = \tilde W_{\text{YM}} +
	 2 \sqrt{3}
	\frac{\tilde E }{\tilde g^2}  \left|\tilde Q_c\right|
	+ \tilde W_{nl}.
\label{3_50}
\end{equation}
Here  $\tilde W_{\text{YM}}$ is the energy of the Yang-Mills field [the integral over the whole volume of the terms in the first square brackets in Eq.~\eqref{1_140}];
$\tilde W_{nl}$ is the energy of the spinor field associated with the nonlinearity of this field
[the integral over the whole volume of the second term appearing in the second square brackets in Eq.~\eqref{1_140}]. Note that for $\tilde E \rightarrow 1$
the total energy of the particlelike object under consideration and its color charge coincide, apart from a constant:
$$	
	\tilde W_t \approx 2  \sqrt{3}
		\frac{\left|\tilde Q_c\right| }{\tilde g^2} \tilde E
		\quad \text{ for } \quad \tilde E \rightarrow 1 .
$$

To ascertain the physical meaning of the parameter $\tilde{\mathfrak{m}}_c$, consider the asymptotic behavior  of the color magnetic field.
Substituting the expression for the potential \eqref{3_30} in the expressions~\eqref{1_70}-\eqref{1_90}, we have
\begin{align}
\tilde{H}^1_i \equiv &	g \lambda_c^2 H^1_i \approx  \frac{\tilde{\mathfrak{m}}_c}{x^3}
	\left\lbrace
		2 \sin \theta \cos \varphi,
		-\cos \theta \cos \varphi ,
		\sin \varphi
	\right\rbrace ,
\label{3_70}\\
\tilde{H}^2_i \equiv & g \lambda_c^2 H^2_i \approx  \frac{\tilde{\mathfrak{m}}_c}{x^3}
	\left\lbrace
		2 \sin \theta \sin \varphi,
		-\cos \theta \sin \varphi ,
		- \cos \varphi
	\right\rbrace ,
\label{3_80}\\	
\tilde{H}^3_i  \equiv & g \lambda_c^2 H^3_i \approx  \frac{\tilde{\mathfrak{m}}_c}{x^3}
	\left\lbrace
		2 \cos \theta,
		\sin \theta ,
		0
	\right\rbrace ,
\label{3_90}
\end{align}
where we have introduced the physical components of the magnetic field vector
$H^a_i = \sqrt{H^a_i H^{a i}}$ (no summing over $a$ and $i$). It is seen from Eq.~\eqref{3_90} that
the asymptotic behavior of the color magnetic field $\vec{H}^3$ coincides exactly with the field of a magnetic dipole in Maxwell's magnetostatic.
This enables us to refer to the parameter  $\tilde{\mathfrak{m}}_c$ as the color magnetic moment.
As one sees from Fig.~\ref{charge_mgn_moment},  the color magnetic moment, as well as  the color charge, has a global minimum, which may be called a magnetic moment gap.

Notice also that, in a color space, the asymptotic behavior of the scalar product of the color vector of the magnetic field $ H^a_i$
and the color vector
$
	l^a = \left\lbrace
		\cos \varphi, \sin \varphi, 0
	\right\rbrace
$
coincides with the magnetic field of a magnetic dipole as well:
$$	g \lambda_c^2 H^a_i l^a \approx \frac{\tilde{\mathfrak{m}}_c}{x^3}
	\left\lbrace
		2 \sin \theta, - \cos \theta, 0
	\right\rbrace ,
$$
but with the change in the direction of the magnetic moment as
$\theta = \theta^\prime + \pi/2$.

\section{Discussion and conclusions}

In the present study, we have demonstrated the existence of charge and magnetic moment gaps
in  SU(3) Yang-Mills theory with a source of a gauge field in the form of a nonlinear spinor field.
We have found particlelike solutions with color electric and magnetic fields belonging to the subgroup
$\text{SU(2)} \times \text{U(1)} \subset \text{SU(3)}$. By the investigation of the asymptotic behavior of such fields, we have shown that the electric field has a Coulomb asymptotic form, and the $\vec{H}^3$ color component of the magnetic field behaves asymptotically as the field of a magnetic dipole.

In the dimensionless form, the regular solutions obtained form a family labeled by the dimensionless spinor frequency~$\tilde E$. For this family, one can construct profiles of the dependence of the color electric charge  $\tilde Q_c$ and of the color magnetic moment  $\tilde{\mathfrak{m}}_c$ on the parameter $\tilde E$, which represents the solutions in this family. As in the case of the energy spectrum, these profiles have global minima, which may be called charge and magnetic moment gaps.

Summarizing the results obtained,
\begin{itemize}
\item The equations of non-Abelian SU(3) Yang-Mills theory with a source in the form of a nonlinear spinor field have regular particlelike solutions with non-Abelian $\text{ SU(2)}\subset \text{SU(3)}$ magnetic and Abelian  $\text{U(1)} \subset \text{SU(3)}$ electric fields.
 \item Asymptotically, the color electric field exhibits the Coulomb behavior;  this enables one to introduce the corresponding charge.
 \item The asymptotic behavior of the color magnetic field is the same as that of a magnetic dipole in Maxwell's electrodynamics;  this enables one to determine the corresponding color magnetic moment.
 \item The profiles of the color charge and magnetic moment, being functions of the system parameter $\tilde E$ (which represents the solution),  have global minima (see Fig.~\ref{charge_mgn_moment}); by analogy with the energy spectrum of these solutions, this enables one to speak about the presence in the system of
 charge and magnetic moment gaps.
\end{itemize}
The reason for the appearance of the mass, charge, and magnetic moment gaps in the theory under consideration is of great importance. In our opinion, the main and the only reason is the presence of a nonlinear spinor field. A mass gap for a system supported by a nonlinear spinor field uncoupled to other fields was first obtained in Refs.~\cite{Finkelstein:1951zz,Finkelstein:1956}. According to our results obtained in the present paper and earlier (see, e.g., Refs.~\cite{Dzhunushaliev:2020qwf,Dzhunushaliev:2021apa}),
the gaps are also present when other fields are involved; that is, this effect is inherent to a nonlinear spinor field, in contrast to, for example, a system involving a nonlinear scalar field where a mass gap is absent.

\section*{Acknowledgments}

The work was supported by the Science Committee of the Ministry of Science and Higher Education of the Republic of Kazakhstan (Grant  No. AP14869140, ``The study of QCD effects in non-QCD theories''). We are also grateful to the Research Group Linkage Programme of the Alexander von Humboldt Foundation for the support of this research.


\begin{thebibliography}{99}

\bibitem{Ayyar:2017qii}
V.~Ayyar and S.~Chandrasekharan,
``Generating a nonperturbative mass gap using Feynman diagrams in an asymptotically free theory,''
Phys. Rev. D \textbf{96},  114506 (2017).

\bibitem{Burikham:2017bkn}
P.~Burikham, T.~Harko, and M.~J.~Lake,
``The QCD mass gap and quark deconfinement scales as mass bounds in strong gravity,''
Eur. Phys. J. C \textbf{77} , 803 (2017).

\bibitem{Faedo:2017fbv}
A.~F.~Faedo, D.~Mateos, D.~Pravos, and J.~G.~Subils,
``Mass Gap without Confinement,''
JHEP \textbf{06}, 153 (2017).

\bibitem{Kojo:2012js}
T.~Kojo and N.~Su,
``The quark mass gap in a magnetic field,''
Phys. Lett. B \textbf{720}, 192 (2013).

\bibitem{heis}
W. Heisenberg, \textit{Introduction to the unified field theory of elementary particles}
(Max-Planck-Institut f\"ur Physik und Astrophysik, Interscience Publishers London, New York, Sydney, 1966).

\bibitem{Volkov:2005kw}
M.~K.~Volkov and A.~E.~Radzhabov,
``The Nambu-Jona-Lasinio model and its development,''
Phys. Usp. \textbf{49}, 551 (2006).

\bibitem{Buballa:2003qv}
M.~Buballa,
``NJL model analysis of quark matter at large density,''
Phys. Rept. \textbf{407}, 205 (2005).

\bibitem{Dzhunushaliev:2020qwf}
V.~Dzhunushaliev, V.~Folomeev, and A.~Serikbolova,
``Monopole solutions in SU(2) Yang-Mills-plus-massive-nonlinear-spinor-field theory,''
Phys. Lett. B \textbf{806}, 135480 (2020).

\bibitem{Dzhunushaliev:2021apa}
V.~Dzhunushaliev, N.~Burtebayev, V.~N.~Folomeev, J.~Kunz, A.~Serikbolova, and A.~Tlemisov,
``Mass gap for a monopole interacting with a nonlinear spinor field,''
Phys. Rev. D \textbf{104},  056010 (2021).

\bibitem{Cheng}
Ta-Pei Cheng and Ling-Fong Li, \textit{
Gauge theory of elementary particle physics}
(Oxford University Press, 1994). 

\bibitem{Li:1982gf}
X.~Z.~Li, K.~L.~Wang, and J.~Z.~Zhang,
Light Spinor Monopole,
Nuovo Cim.\ A {\bf 75}, 87 (1983).

\bibitem{Li:1985gf}
K.~L.~Wang and J.~Z.~Zhang,
``The Problem of Existence for the Fermion-Dyon Selfconsistent Coupling System in a SU(2) Gauge Model,''
Nuovo Cim.\ A {\bf 86}, 32 (1985).

\bibitem{Finkelstein:1951zz}
R.~Finkelstein, R.~LeLevier, and M.~Ruderman,
``Nonlinear Spinor Fields,''
Phys.\ Rev.\  {\bf 83}, 326 (1951).

\bibitem{Finkelstein:1956}
R. Finkelstein, C. Fronsdal, and P. Kaus,
``Nonlinear Spinor Field,''
Phys.\ Rev.\  {\bf 103}, 1571 (1956).

\end{thebibliography}
\end{document}